\documentclass[twocolumn]{aastex631}

\graphicspath{{./}{}}
\usepackage{amsmath}
\usepackage{comment}
\usepackage{wasysym}
\usepackage{booktabs}

\begin{document}

\title{The Narrow Formation Pathway of Hot Saturns: Constraints on Initial Planetary Properties}

\correspondingauthor{Sheng Jin}
\email{jins@ahnu.edu.cn}

\author{Minghao Xie}
\affiliation{Department of Physics, Anhui Normal University, Wuhu, Anhui 241002, China}

\author[0000-0002-9063-5987]{Sheng Jin}
\affiliation{Department of Physics, Anhui Normal University, Wuhu, Anhui 241002, China}

\author[0000-0001-9424-3721]{Dong-Hong Wu}
\affiliation{Department of Physics, Anhui Normal University, Wuhu, Anhui 241002, China}

\begin{abstract}

The observed exoplanet population exhibits a scarcity of short-period Saturn-mass planets, a phenomenon referred to as the ``hot Saturn desert".
This observational scarcity can be utilized to validate the theories regarding the formation and evolution of gas planets.
In this study, we conduct large-scale numerical simulations to explore how the initial conditions of gas planets orbiting solar-type and M-dwarf stars influence their evolutionary trajectories in the semi-major axis versus planetary radius ($a$-$R$)  parameter space.
We generate a synthetic population of 10,000 short-period gaseous planets by systematically varying their initial planetary masses ($M_{\rm p}$), initial planetary luminosities ($L_{\rm p}$), initial core mass fractions ($f_{\rm core}$), and semi-major axis ($a$).
Furthermore, we assume these gaseous planets have ceased orbital migration and model their long-term thermal evolution, taking into account the impacts of atmospheric evaporation.
Our results show that the initial mass, $L_{\rm p}$, and $f_{\rm core}$ are the dominant factors controlling radius evolution for short-period gas planets.
The key to survival as a hot Saturn analogue appears to be having just the right combination of properties after gas disk dissipation: an $M_{\rm p}$ below 0.5 Jupiter Mass ($M_{\rm Jup}$), a substantial $f_{\rm core}$ of $\geq$30\%, and relatively low $L_{\rm p}$ on the order of $10^{-6}$ solar luminosity ($L_{\odot}$) or less.
The survival criteria for hot Saturn analogs align with theoretically unfavorable initial conditions of gas planets formed via core accretion scenario, naturally explaining the observed boundaries of the hot Saturn desert.

\end{abstract}

\section{Introduction}
\label{sec:intro}

The exoplanet population exhibits a  notable scarcity of short-period planets with Saturn-like masses and radii, creating a distinct ``hot-Saturn desert" in the $a$-$R$ parameter space \citep{Zhu2021}.
While examples exist (e.g., TOI-1194 b, a planet with a radius of $0.767^{+0.045}_{-0.041} R_{\rm Jup}$ and a minimum mass of $0.456^{+0.055}_{-0.051} M_{\rm Jup}$; \citet{Wang2024}),
their occurrence rate remains substantially lower than both hot Jupiters and close-in terrestrial planets.

From a theoretical perspective, the absence of hot Saturns can be replicated and understood through planetary population synthesis models that are grounded in the core accretion scenario \citep{Ida2004,Ida2005,Mordasini2012a,Mordasini2012b}.
On one hand, when a protoplanet reaches the crossover mass, defined as the point where its solid core mass equals its gaseous envelope mass, typically in the range of 10-30 $M_{\oplus}$, it undergoes runaway gas accretion and rapidly evolves into a gas giant \citep{Pollack1996}.
On the other hand, gas giants are more likely to form in the outer regions of the protoplanetary disk, where there is a greater abundance of material. 
These giant planets can arrive short-period orbits only if they are sufficiently large to trigger Type II migration, as outlined in the gas disk migration scenario \citep{Lin1996}.
Taking these two factors into account, it is challenging for Saturn-like planets to reside in short-period orbits. Even if such planets initially reach close-in orbits via disk migration, dynamical instabilities in the post-disc phase may further sculpt the final architecture of planetary systems, potentially removing planets that formed near the desert boundary \citep{Wu2023,He2024}.
Furthermore, high-eccentricity migration and atmospheric evaporation processes also play a significant role in sculpting the boundaries of the hot Saturn desert \citep{Matsakos2016,Owen2018}. 
The synergy of these mechanisms leads to the observed scarcity of hot Saturn exoplanets.

Although theoretical models based on the core accretion scenario provide a conceptual explanation for the desert of hot Saturn exoplanets, they do not quantitatively specify which types of planets are unfavorable to form.
Because formation theories typically produce all theoretically viable outcomes, rather than predicting which specific outcomes are prohibited.
From an observational point of view, the hot Saturn desert manifests as a distinct forbidden region in  $a$-$R$ space, where planets ending up in this region are considered unfavorable.
This desert is a statistical feature that emerges from the exoplanet demographics, independent of specific planet formation models.
By examining the shape of the hot-Saturn desert, we can backtrack to the early stages of planet formation and quantitatively identify which types of planets are unlikely to form, namely those that would ultimately reside in the desert region.

Assuming that there is a virtual population of short-period planets that, after undergoing long-term thermal evolution, could eventually occupy the region now known as the hot Saturn desert, then the initial physical parameters of these hypothetical planets would offer crucial constraints on planet formation theories.
Given that the evolved counterparts—hot Saturn analogues—of this hypothetical population are rarely observed in reality, understanding why planets with such initial parameters fail to form in short-period orbits can provide valuable insights into the early stages of planet formation when a gas disk is still present.

In this study, we aim to find out the initial physical parameters of the hypothetical population of short-period gaseous planets that are unlikely to exist, as inferred from the shape of the hot Saturn desert.
The observed hot Saturn desert represents a distribution feature at a very late stage, after planets have formed, migrated, and undergone long periods of thermal evolution.
Hence, we trace back to the early stage when the gas disks had just dissipated and the disk-based migration had just ceased, in order to identify the unfavorable initial planetary parameters.
We bypass the constraints of theoretical planet formation models to generate a large number of short-period gaseous planets with a wide range of initial physical parameters.
We then track the evolutionary trajectories of these synthetic gaseous planets and determine their final positions in the $a$-$R$ space.
Our results show that the evolution of short-period gaseous exoplanets is strongly linked to atmospheric evaporation, with three primary initial parameters playing a crucial role: $M_{\rm p}$, $L_{\rm p}$, and $f_{\rm core}$.
We also find that a group of hot Saturn analogues have ultimately ended up in the hot Saturn desert region.
Notably, the initial physical parameters of these  hot Saturn analogues, which are essentially the forbidden parameters given the existence of the observed hot Saturn desert,  align with the unfavorable initial  parameters predicted by the core accretion scenario.

This paper is organized as follows. Section \ref{sec:model} describes our planet grids and our planetary structure and evolution model.
In section \ref{sec:results}, we present the simulation outcomes around solar-type stars, analyze the influence of varying initial planetary parameters, and examine the specific case of planets orbiting M-dwarf stars.
Section \ref{sec:discussion}  provides a brief summary and discussion.

\section{MODEL}
\label{sec:model}

\subsection{Grid of synthetic gas planets} 

Gaseous planets primarily complete their formation while their host star still retains a protoplanetary disk.
During this stage, the  orbital semi-major axis distribution of gaseous planets is determined by two main factors: the amount of available material at different locations within the disk, and the degree of orbital migration resulting from the interaction between a planet and the gas disk.
Here, we focus on the physical parameters of gaseous planets immediately following the initial formation stage, specifically when the protoplanetary gas disks have just dissipated and the planets have ceased mass accretion and orbital migration. 
In this context, the term ``initial planetary parameters" used in this paper refers to the planetary parameters just after disk dissipation.

This study aims to determine the types of planets that 
can end up in the hot Saturn desert region after long-term evolution, rather than focusing on whether such planets can form under any specific formation scenario.
Therefore, we expand the ranges of initial planetary parameters to encompass as many types of simulated short-period gaseous planets as possible.
We focus on four key parameters of a gaseous planet, initial planetary mass $M_\mathrm{p}$, initial  planetary luminosity $L_\mathrm{p}$, initial core mass fraction $f_\mathrm{core}$, and the planetary orbital semi-major axis $a$.
Given that this study aims to identify the early-stage counterparts of hot Saturn analogues, we set the initial masses of our simulated planets within the range of 0.3 to 1.2 $M_{\rm Jup}$.
The $a$ is set between 0.03 and 0.1 AU.
The initial $f_\mathrm{core}$ is set between 0.1\% and 50\%.
The initial $L_\mathrm{p}$ is set between  $8.7 \times 10^{-10}$ and $1.7 \times 10^{-5}$ $L_{\odot}$.
Table \ref{tab:1} provides the details of the settings for this four-dimensional parameter grid.
We then generate 10,000 synthetic gaseous planets using various combinations of these four parameters.

\begin{table}[]
\caption{A four-dimensional parameter grid of initial conditions for 10,000 synthetic gas giants.}
\label{tab:1}
\begin{tabular}{cccc} \hline
$M_{\rm p}$ & $a$& $f_\mathrm{core}$& $L_{\rm p}$ \\ 
  ($M_{\rm Jup}$)&  ($AU$)&  &  (\( L_{\odot} \))\\ \hline
  0.3&   0.03&       0.1\%&   $8.7 \times 10^{-10}$\\
  0.4&   0.035&       0.3\%&   $2.6 \times 10^{-9}$\\
  0.5&   0.04&       1\%&   $7.8 \times 10^{-9}$\\
  0.6&   0.045&       3\%&   $2.3 \times 10^{-8}$\\
  0.7&   0.05&       10\%&   $7.0 \times 10^{-8}$\\
  0.8&   0.06&       15\%&   $2.1 \times 10^{-7}$\\
  0.9&   0.07&       20\%&   $6.3 \times 10^{-7}$\\
  1.0&   0.08&       30\%&   $1.9 \times 10^{-6}$\\
  1.1&   0.09&       40\%&   $5.7 \times 10^{-6}$\\
  1.2&   0.1&       50\%&   $1.7 \times 10^{-5}$\\ \hline
\end{tabular}
\end{table}

\subsection{Planetary Structure Model}

Our simulations utilize a two-layer planetary structure model, featuring a solid planetary core enveloped by a hydrogen-helium (H/He) envelope.
Regardless of the core mass, every planetary core follows the same compositional template: 5.1\% of its mass as ice, while the other 94.9\% precisely mirrors Earth’s silicate-iron distribution (2/3 silicate and 1/3 iron by mass).
The core mass remains fixed throughout the evolutionary process.
The instantaneous core radius is determined based on three parameters: core mass, ice mass fraction, and the pressure at the envelope-core boundary \citep{Mordasini2012b}.
We assume spherical symmetry in the planetary H/He envelope and calculate its structure by solving a set of one-dimensional hydrostatic equilibrium equations \citep{Jin2014}.

The temperature gradient within the gaseous envelope is influenced by both the optical depth and the heat transfer mechanism (whether convective or radiative) at each layer of the envelope.
We partition the gaseous envelope into two segments: the atmosphere, where the majority of the stellar irradiation is absorbed, and the underlying layer referred to as the envelope.
If an atmospheric layer is convectively stable,
we adopt the globally averaged temperature profile
from Equation (49) in the semi-grey model proposed by \citet{Guillot2010}:

\begin{equation}
\begin{aligned}
T^4 &= \frac{3T_{\text{int}}^4}{4} \left( \frac{2}{3} + \tau \right) \\
&\quad + \frac{3T_{\text{eq}}^4}{4} \left\{ \frac{2}{3} + \frac{2}{3\gamma} \left[ 1 + \left( \frac{\gamma \tau}{2} - 1 \right) e^{-\gamma \tau} \right] \right. \\
&\quad \left. + \frac{2\gamma}{3} \left( 1 - \frac{\tau^2}{2} \right) E_2(\gamma \tau) \right\}
\end{aligned}
\label{2bdglobal}
\end{equation}
where \(T_{\text{int}}\) is the planet's intrinsic temperature, \(T_{\text{eq}}\) is the planet's equilibrium temperature, \(\tau\) is the optical depth, and \(\gamma\) is the ratio of visible opacity to thermal opacity. 

The transition from atmospheric to envelope regimes occurs when the visible-wavelength optical depth $\tau_{\rm v}$ substantially exceeds unity.
According to the $\gamma$ defined in the semi-grey model,
we establish $\tau \gg 1/(\sqrt{3}\gamma)$ at the transition \citep{Rogers2011}.
Since most of the starlight is absorbed at pressures less than 10 bar
\citep{Guillot2002}, we set the atmosphere/envelope boundary
at $\tau = 100/(\sqrt{3}\gamma)$, which corresponds to a pressure of approximately 10 bar.
If the envelope at $\tau > 100/(\sqrt{3}\gamma)$ is convectively stable,
the radiative temperature gradient is calculated using the diffusion
approximation, which take into account only the planet's intrinsic luminosity:
\begin{equation}
  \frac{{\rm d}T}{{\rm d}r}=-\frac{3\kappa_{\rm th}\rho L}{64\pi\sigma_{B}T^3 R^2}
\end{equation}

Convective instability arises when the radiative temperature gradient surpasses the adiabatic gradient, leading to the application of the adiabatic temperature profile throughout the unstable region:
\begin{equation}
  \frac{{\rm d}T}{{\rm d}r}=\frac{T}{P}\frac{{\rm d}P}{{\rm d}r}\left(\frac{{\rm ln}T}{{\rm ln}P}\right)_{\rm ad}
\end{equation}
where the adiabatic temperature gradient is established using the equation of state from \citet{Saumon1995}. This convective adjustment is similarly applied in the atmosphere; however, we do not allow for detached convective zones in that region.

\subsection{Planetary Evolution and Atmospheric evaporation}
\label{model:2.3}

After the dissipation of protoplanetary disks, the processes of disk-driven migration and mass accretion (including planetesimals and nebular gas) cease, marking the beginning of the planetary evolutionary phase. 
For short-period planets experiencing substantial stellar irradiation, atmospheric evaporation plays a crucial role in shaping the outcomes of a planet's billion-year evolution \citep{Lammer2003,Baraffe2004,Lopez2012,Owen2013,Jin2014,Luger2015,Chen2016,Jackson2017,Kubyshkina2020,Ketzer2023,Guo2024}, especially in the emergence of the hot Saturn desert region \citep{Owen2018}.
By conducting numerical simulations of planetary thermal evolution coupled with atmospheric evaporation, we can constrain important planetary properties both at present and in the past \citep{Valencia2010,Jackson2010,Jackson2016,Kubyshkina2019,Kubyshkina2022,Ketzer2024}.

We employ a combined thermodynamic evolution and atmospheric evaporation model to simulate the radius evolution of short-period gas giants, with comprehensive details provided in \citet{Mordasini2012b,Jin2014}.
For clarity, we summarize here the key aspects of our atmospheric evaporation model.
Our evaporation model  incorporates the influence of both X-ray and extreme ultraviolet (EUV) radiation emitted by the host star, dividing the atmospheric escape process into X-ray-dominated and EUV-dominated regimes \citep{Murray2009,Owen2012}.
X-ray photons have smaller interaction cross-section and hence penetrate deeper in 
 the planetary atmosphere compared with EUV photons.
Following \citet{Owen2012}, we define the evaporation process as X-ray-driven when the X-ray-powered flow achieves the sonic condition before meeting the hydrogen ionization front generated by EUV radiation.
This typically occurs during early planetary evolutionary stages when young stars exhibit strong X-ray emission \citep{Ribas2005,Jackson2012,Jin2014}.
In this X-ray-driven regime, we calculate the atmospheric mass-loss rate using an energy-limited model proposed by \citep{Owen2012}:
\begin{equation}
  \dot{m}=\epsilon\frac{\displaystyle{16 \pi F R^3_{\rm p}}}{\displaystyle{3 G M_{\rm p} K(\xi)}}
\label{Mdotxray}
\end{equation}
where the $\epsilon$ is the heating efficiency that is in the range of 0.1-0.25 \citep{Lammer2009,Jackson2012},
$M_{\rm p}$ is the planetary mass,
$R_{\rm p}$ is the planet's radius at
the optical depth $\tau = 2/3$ in the thermal wavelengths.
($\tau$ is calculated using the grain-free Rosseland mean opacity
$\kappa_{\rm th}$ from \citet{Bell1994} and \citet{Freedman2008}),
$F$ is the X-ray flux in the wavelength range from
1 to 20 ${\rm \AA}$ from \citet{Ribas2005},
$\xi=R_{\rm roche}/{R_{\rm p}}$, and
\begin{equation}
K(\xi)=1-\frac{3}{2\xi}+\frac{1}{2\xi^3}
\end{equation}
accounts for the enhanced mass-loss rate by a factor of
$1/K(\xi)$ because the Roche lobe of a close-in planet can be
close to the planet's surface \citep{Erkaev2007}. 

EUV photons ionizes upper atmospheric hydrogen through photoelectric effect.
The balance between photoionization and recombination processes yields an equilibrium temperature of $\sim$ $10^4$ K, corresponding to the peak in H recombination cooling \citep{Dalgarno1972}.
When the X-ray-driven flow fails to reach sonic velocity before encountering this hydrogen ionization front, the atmospheric evaporation shifts to the EUV-driven regime.
The resulting EUV-driven evaporation can be categorized into two sub-regimes based on the magnitude of EUV irradiation \citep{Murray2009}.
Under strong EUV irradiation, radiative cooling dissipates much of the incident energy, causing the mass-loss rate to follow a weaker $\dot{M} \propto F_{EUV}^{0.6}$ relationship \citep{Murray2009}.
This behavior cannot be captured by energy-limited approaches and is referred to as the radiation-recombination-limited regime. 
In this regime, the mass-loss rate is limited by the balance between radiative cooling and recombination processes, rather than by the energy input from stellar irradiation.
In accordance with  \citet{Murray2009}, we compute the atmospheric mass-loss rate under intense EUV irradiation using:
\begin{equation}
\dot{M}_{\rm rr-lim} \sim 4 \pi \rho_{\rm s} c_{\rm s} r_{\rm s}^2
\end{equation}
where $c_{\rm s} = [kT/(m_{\rm H}/2)]^{1/2}$ is the isothermal sound speed,
$T = 10^4$ K is the temperature at the sonic point,
$r_{\rm s} = GM_{\rm p}/(2c_{\rm s}^2)$ is the sonic point where
the mass flow escapes the planet at the sound speed, and $\rho_{\rm s}$ is
the flow density at the sonic point.

 At low EUV fluxes ($<10^{4}$ erg cm$^{-2}$ s$^{-1}$), we recover the energy-limited approximation to estimate the EUV-driven mass-loss rates \citep[e.g.,][]{Watson1981,Murray2009}:
\begin{equation}
  \dot{M}_{\rm e-lim}=\epsilon\frac{\displaystyle{\pi F_{\rm EUV} R^3_{\rm base}}}{\displaystyle{G M_{\rm p} }}
  \label{Mdotxray}
\end{equation}
where $\epsilon$ is the heating efficiency (taken as 0.3 as found in \citet{Murray2009}), $F_{\rm EUV}$ the EUV flux at the position of the planet, $R_{\rm base}$ the radius of the photoionization base estimated as \citet{Murray2009}, $M_{\rm p}$ the planet mass and $G$ the gravitational constant.

Note that the actual atmospheric escape behavior and heating efficiencies for individual planets vary significantly with planetary parameters (mass, radius, atmospheric composition) and stellar irradiation, all of which are time-dependent \citep{Yelle2004,Tian2005,Owen2012,Owen2016}. However, as shown in \citet{Jin2014}, the statistical signatures of atmospheric evaporation across planetary populations remain robust to moderate variations in the intensity of evaporation.

\section{Results}
\label{sec:results}

\subsection{Emergence of Hot Saturns}

\begin{figure*}
\centering
\includegraphics[width=0.75\textwidth]{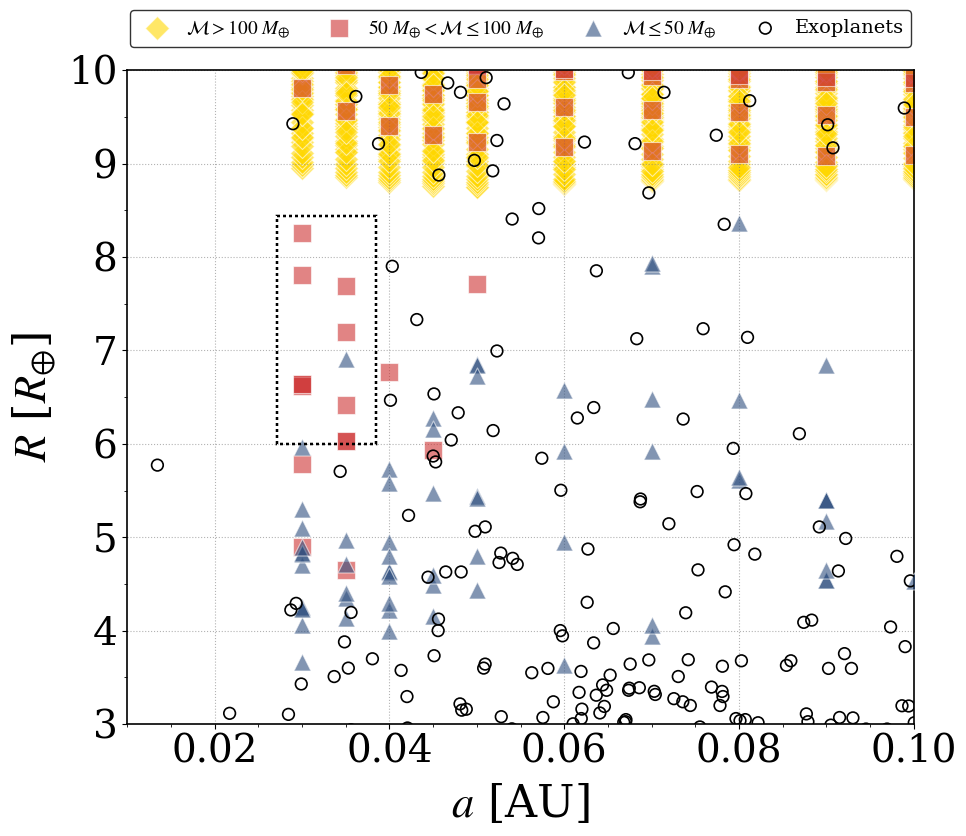}
\caption{
The final $a$-$R$ distribution  of our synthetic short-period gaseous planets at 1 Gyr is plotted alongside the known exoplanet population. 
Given our focus on planets that can end up in the hot Saturn desert region, the plot only includes planets with radii between 3 and 10 $R_{\oplus}$. 
The colored points represent our synthetic planets within different mass ranges after 1 Gyr of evolution, while the circles indicate the observed exoplanets.
The dotted box highlights the region where hot Saturn analogues are formed.
}
\label{fig:1}
\end{figure*}

\begin{table*}[]
\centering
\caption{Initial and final properties of hot Saturn analogues emerging from 10,000 synthetic planets orbiting a solar-type star.}
\label{tab:2}
\begin{tabular}{ccccccccccc} 
\toprule
 & & \multicolumn{5}{c}{Initial} & \multicolumn{4}{c}{Final} \\
\cmidrule(lr){3-7} \cmidrule(lr){8-11}
Index & $a$  &  $M_{\rm p}$  &  $f_\mathrm{core}$ &  $L_{\rm p}$  &  $S_\mathrm{inner}$$^{(1)}$ & $S_\mathrm{outer}$$^{(2)}$ &  $M_{\rm p}$  & $f_\mathrm{core}$ &  $L_{\rm p}$  &  $R_{\rm p}$  \\
 &  (AU) &  ($\mathrm M_{\rm Jup}$) &  & ($\mathrm L_{\odot}$)  & ($K_\mathrm{B}/baryon$) &($K_\mathrm{B}/baryon$) & ($\mathrm M_{\rm Jup}$) &  &  ($\mathrm L_{\odot}$) & ($\mathrm R_{\oplus}$) \\ 
 \midrule
 01&  0.03&  0.5&  40\%&  $1.9 \times 10^{-6}$ & 11.00 &11.32 &  0.23&  88.5\%&  $1.93 \times 10^{-10}$& 6.64 \\
 02&  0.03&  0.5&  50\%&  $1.9 \times 10^{-6}$ & 10.93 &11.07 &  0.31&  80.4\%&  $5.46\times 10^{-10}$& 7.81 \\
 03&  0.035&  0.5&  30\%&  $1.9 \times 10^{-6}$ & 10.75 &10.99 &  0.16&  91.0\%&  $8.77 \times 10^{-11}$& 6.03 \\
 04&  0.035&  0.5&  40\%&  $5.7 \times 10^{-6}$ & 10.96 &10.98 &  0.25&  80.5\%&  $3.67 \times 10^{-10}$& 7.69  \\
 05&  0.03&  0.4&  50\%&  $6.3 \times 10^{-7}$ & 10.68 &11.17 &  0.23&  88.5\%&  $1.88 \times 10^{-10}$& 6.62 \\
 06&  0.03&  0.4&  50\%&  $1.9 \times 10^{-6}$ & 11.14 &11.61 &  0.23&  88.4\%&  $1.95 \times 10^{-10}$& 6.64 \\
 07&  0.035&  0.4&  40\%&  $1.9 \times 10^{-6}$ & 10.89 &11.26 &  0.18&  88.9\%&  $1.25\times 10^{-10}$& 6.42  \\
 08&  0.035&  0.4&  50\%&  $5.7 \times 10^{-6}$ & 11.28 &11.44 &  0.24&  83.7\%&  $2.84 \times 10^{-10}$& 7.20 \\
 09&  0.03&  0.3&  50\%&  $2.1 \times 10^{-7}$ & 9.36 &10.77 &  0.18&  81.2\%&  $2.58 \times 10^{-10}$& 8.25  \\
 10&  0.035&  0.3&  30\%&  $7.0 \times 10^{-8}$ & 8.75 &10.51 &  0.10&  89.0\%&  $6.87 \times 10^{-11}$& 6.91  \\
 11&  0.035&  0.3&  50\%&  $6.3 \times 10^{-7}$ & 10.61 &11.24 &  0.16&  91.0\%&  $8.79 \times 10^{-11}$& 6.03  \\ \hline
 \multicolumn{11}{l} { \footnotesize (1) $S_\mathrm{inner}$ is the entropy at the core-envelope boundary.} \\
  \multicolumn{11}{l} { \footnotesize (2) $S_\mathrm{outer}$ is the entropy at the planetary radius of $\tau = 2/3$.} \\
 \bottomrule
\end{tabular}
\end{table*}

Using the model outlined in section \ref{sec:model}, we simulate the thermal evolution over 1 billion years (Gyr) for all 10,000 synthetic short-period gaseous planets, incorporating the effects of hydrodynamic atmospheric evaporation.
Our simulations show that, under most initial conditions, synthetic planets lose a moderate amount of atmosphere and settle into the hot Jupiter domain. 
However, a subset of initial conditions produced planets that instead reside in the hot Saturn desert, appearing as an undiscovered population in the $a$-$R$ diagram when compared to observed exoplanets.

In Figure \ref{fig:1}, we present the final $a$-$R$ distribution of our synthetic short-period gaseous planets with a final radius between 3 and 10 Earth radii ($R_{\oplus}$).
Considering our focus on planets that may end up in the hot Saturn desert zone, the graph only shows planets with radii ranging from 3 to 10 $R_{\oplus}$.
It is evident that the majority of our synthetic gaseous planets have radii larger than 8.5 $R_{\oplus}$,  as indicated by the clustering of colored points in the upper portion of the figure. 
 These large gaseous planets typically have masses greater than 150 Earth masses ($M_{\oplus}$) and are situated in the lower part of the hot Jupiter region.
Interestingly, we observe that in the hot Saturn desert region, where no exoplanets are currently confirmed, a group of synthetic planets has formed. 
These planets have planetary radii between 6 and 8.5  $R_{\oplus}$ and a semi-major axis smaller than 0.04 AU.
In this paper, we refer to these medium-size, short-period gaseous giant planets as hot Saturn analogues.

\begin{figure*}
\centering
\includegraphics[width=1.005\textwidth]{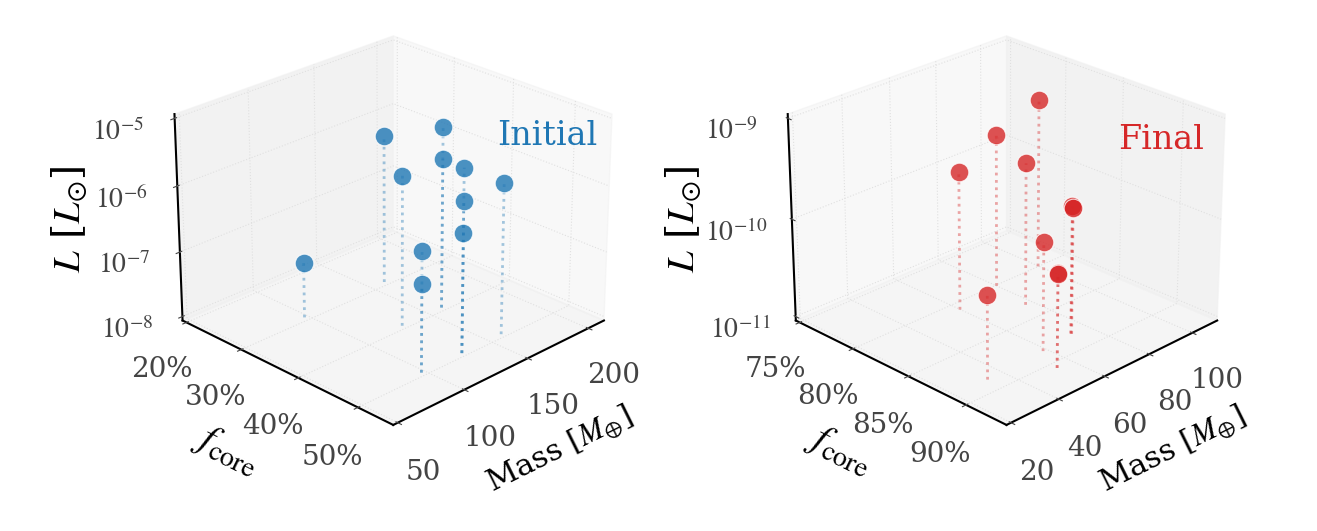}
\caption{The initial (left) and final (right) values of planetary  $L_\mathrm{p}$, $f_\mathrm{core}$, and total mass for the 11 hot Saturn analogues listed in Table \ref{tab:2}.}
\label{fig:2}
\end{figure*}

Table \ref{tab:2} lists  the initial and final planetary parameters of the 11 hot Saturn analogues that emerged in our synthetic planet population.
All of these hot Saturn analogues have a final $f_{\rm core}$ exceeding 80\%, which suggests that they have undergone substantial mass loss due to atmospheric evaporation.
For example, consider the hot Saturn labeled  as 01 in in Table \ref{tab:2}.
Initially, it has a planetary mass of 0.5 $M_{\rm Jup}$, a $f_{\rm core}$ of 40\%, a $L_{\rm p}$ of $1.9\times10^{-6}$ $L_{\odot}$, and an orbital distance of 0.03 AU.
Ultimately,  this planet has lost $\sim$ 91.0\% of its initial envelope, resulting in a final $f_{\rm core}$ of $\sim$ 88.5\%.
The remaining planetary mass is $\sim$ 0.23 $M_{\rm Jup}$, with a relatively low $L_{\rm p}$  of $1.93\times10^{-10}$ $L_{\odot}$.
Its orbital distance remains at 0.03 AU, as planetary migration is not included in our planetary evolution model.

The hot Saturn analogues listed in Table \ref{tab:2}  exhibit several common features in their initial conditions. 
These characteristics have led them to lose the majority of their initial envelopes and ultimately reside in the hot Saturn desert region.
First, their masses are relatively small, all less than 0.5 $M_{\rm Jup}$. 
Second, they have massive planetary cores, with a $f_{\rm core}$  exceeding 30\%. 
Finally, they exhibit only modest initial luminosity, typically with  $L_{\rm p}$  values around or below $10^{-6}$ $L_{\odot}$.
Table \ref{tab:2} also lists the initial (post-formation) entropies of these hot Saturn analogues, which typically range between 10 and 11 $K_\mathrm{B}/baryon$  for a 0.5  $M_{\rm Jup}$ planet.

Apparently, the formation of hot Saturns depends critically on atmospheric evaporation strength. The process must be vigorous enough to substantially strip the initial gaseous envelope, yet not so extreme as to completely remove it, which would leave only a bare rocky core.
This precisely constrained atmospheric evaporation strength can only result from the coupled effects of the initial planetary  $L_\mathrm{p}$, $f_\mathrm{core}$, and total planetary mass. 
These parameters must fall within a narrow range to maintain the delicate balance required for optimal atmospheric escape, leading to the formation of hot Saturns.
Figure \ref{fig:2} plot the initial and final values of planetary  $L_\mathrm{p}$, $f_\mathrm{core}$, and total mass for the 11 hot Saturn analogues listed in Table \ref{tab:2}.
It shows that 11 hot Saturn analogues are low-mass gas giants ($\leq$ 0.5 $M_{\rm Jup}$) featuring substantial rocky cores and moderate initial luminosities at the onset of their evolution, immediately following the dissipation of the gas disk.
At the end of their 1 Gyr evolution, they become evaporation remnants with a final $f_\mathrm{core}$  ranging between 80\% and 90\%.
In the following sections, we will conduct a detailed analysis of how the initial $L_{\rm p}$ and $f_{\rm core}$ influence planetary evolution by systematically varying the initial properties of planet 01 from Table \ref{tab:2}.

\subsection{Impact of initial $L_{\rm p}$}
\label{sec:3.2}

Our synthetic short-period gaseous planets were generated using a manual grid, differing from standard planet population synthesis approaches that typically employ planetary formation simulations \citep{Ida2004,Mordasini2012a,Mordasini2012b}.
This explains why we are able to form hot Saturns in our study, whereas planet population synthesis models do not.
In this section, we will delve into the reasons behind the formation of hot Saturn analogues by examining how different initial values impact planetary evolution outcomes. 

As indicated in Table \ref{tab:2}, the large final $f_{\rm core}$ of all 11 planets that end up in the hot Saturn desert suggests that the most critical factor in the formation of hot Saturn is the substantial mass loss  due to atmospheric evaporation.
Consequently,  a potential hot Saturn planet must have a relatively low initial planetary mass  to facilitate significant mass loss through atmospheric evaporation at a relatively low gravitational binding energy. 
In fact, all 11 hot Saturns formed in our simulation have an initial planetary mass between 0.3 and 0.5  $M_{\rm Jup}$ and are located within 0.035 AU, where they are subjected to intense stellar radiation.

\begin{figure*}
\centering
\includegraphics[width=1.005\textwidth]{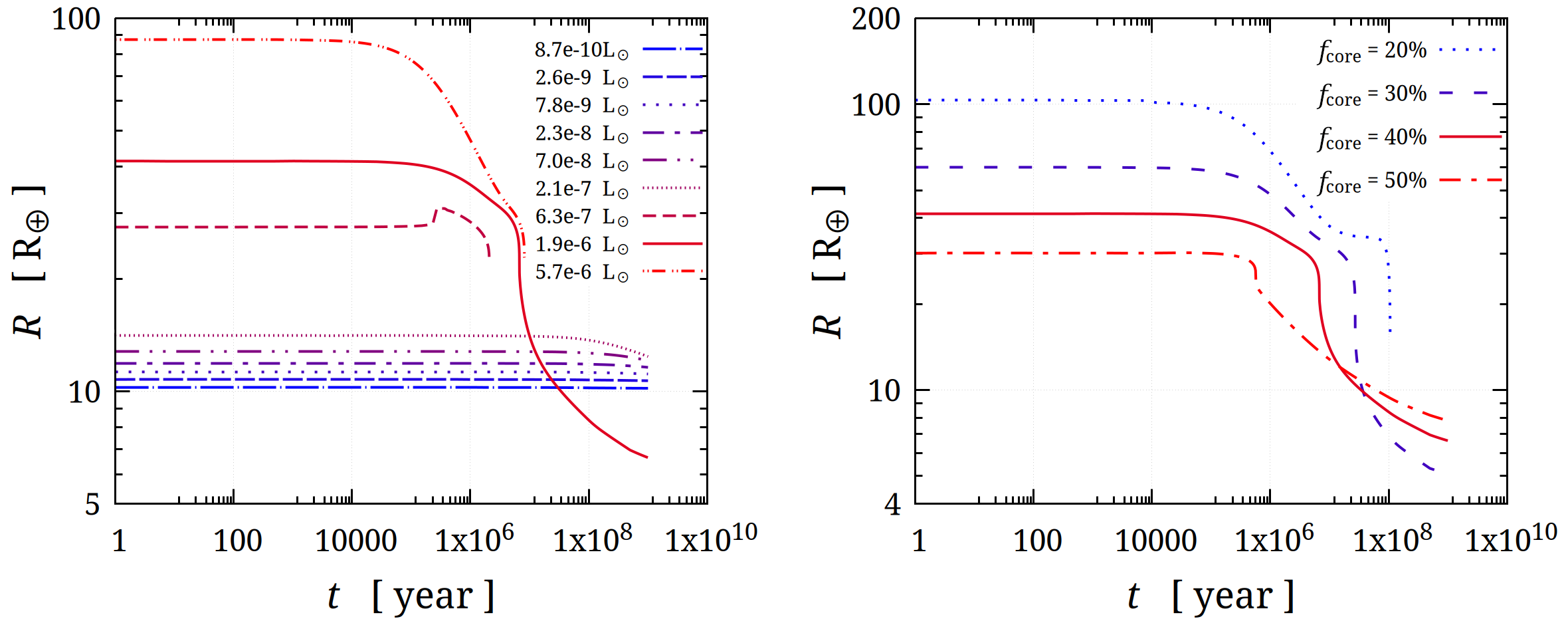}
\caption{The left panel shows the evolutionary trajectories of a 0.5 $M_{\rm Jup}$ planet with a substantial $f_\mathrm{core}$ of $40$\%, but at varying initial $L_{\rm p}$. 
The right panel shows the evolutionary trajectories of a 0.5 $M_{\rm Jup}$ planet with an initial $L_{\rm p}$ of 1.9 $\times$ $10^{-6}$ $L_\odot$,  but with different $f_\mathrm{core}$.}
\label{fig:3}
\end{figure*}

A crucial factor influencing the thermal evolution of gaseous planets is their initial intrinsic luminosity. 
Taking the typical hot Saturn analogue (indexed as 01 in Table \ref{tab:2}) as a reference, 
we selected 10 short-period gaseous planets from our synthetic population of 10,000 that differ from planet 01 solely in their initial $L_{\rm p}$.
This comparison group shares identical properties with planet 01: 0.5  $M_{\rm Jup}$ mass, 40\% $f_\mathrm{core}$, and 0.3 AU orbital distance, while systematically varying only in initial luminosity.
The  evolutionary trajectories of these 10 planets are shown in  the left panel of Figure \ref{fig:3}.

For planets with an extremely high initial luminosity of $1.7\times10^{-5}$ $L_{\odot}$, the high internal entropy prevents them from maintaining a stable planetary structure. As a result, our simulation terminates early, and no evolutionary trajectory is shown for these planets in the left panel of Figure \ref{fig:3}.
For planets with initial luminosities of $5.7\times10^{-6}$ $L_{\odot}$ and $6.3\times10^{-7}$ $L_{\odot}$, their large initial luminosities lead to substantial initial planetary radii. Consequently, they receive significant amounts of incoming stellar flux, which drives intense mass-loss processes. 
Within 10 million years, these two planets lose their entire gaseous envelopes and become rocky cores  \citep{Baraffe2004}.
For planets with initial luminosities less than or equal to $2.1\times10^{-7}$ $L_{\odot}$, they experience mild mass-loss processes.
After 1 billion years of evolution, they retain most of their initial gaseous envelope and evolve into hot Jupiter analogues.
Only the planet with an initial luminosity of $1.9\times10^{-6}$ $L_{\odot}$ undergoes the optimal amount of mass loss.
It loses most of its initial envelope, resulting in a final structure with a large rocky core and a small remaining gaseous envelope. This planet evolves into a hot Saturn analogue.
The results indicate that the evolutionary outcome of a short-period, big-core, Saturn-mass planet is highly sensitive to its initial luminosity—a dependence similarly observed in the long-term evolution of short-period low-mass planets \citep{Kubyshkina2021}.

It is important to note that while the initial planetary radius generally increases with initial planetary luminosity, the mass-loss rate due to atmospheric evaporation is not a monotonic function of initial luminosity.
This is because the relative distance between the surface of a close-in planet and its Roche lobe is not a monotonic function of the planetary radius, which in turn affects the correction factor $K(\xi)$ of the mass-loss rate \citep{Erkaev2007}.
Since we employ semi-analytical equations to calculate the mass-loss rate, the results presented in Figure \ref{fig:3} offer only a qualitative understanding of the relationship between a planet's initial luminosity and its final evolutionary outcome, rather than a strictly quantitative correlation.
Essentially, for a gaseous planet to evolve into a hot Saturn analogue, the amount of atmospheric evaporation must be sufficient to strip away most of its initial atmosphere, but not so strong that it removes the entire gaseous envelope, leaving behind a bare rocky core.
Only within a specific and relatively narrow range of initial luminosities can a gaseous planet ultimately evolve into a hot Saturn analogue.

\subsection{Impact of initial $f_{\rm core}$}
\label{sec:3.3}

Another key factor governing the thermal evolution of a short-period gaseous planet is the initial core mass fraction $f_{\rm core}$.
A large rocky core is crucial for a gaseous planet to maintain  structural stability, particularly during the early stages of planetary evolution when intrinsic luminosity is high. 
Because a substantial core enhances gravitational binding, enabling the planet to retain a significant gaseous envelope despite intense stellar irradiation and atmospheric evaporation processes.

To better understand the influence of the initial $f_{\rm core}$ of a short-period gaseous planet on its radius evolution, we have collected from our 10,000 synthetic planets the evolutionary trajectories of the 10 planets that have only a different initial $f_{\rm core}$ compared with planet 01 in Table \ref{tab:2}. 
Specifically, all these 10 planets are 0.5 $M_{\rm Jup}$ planets with an initial luminosity of $1.9\times10^{-6}$ $L_{\odot}$, located at 0.3 AU, but with varying initial $f_{\rm core}$. 

Given the high initial luminosity of these 10 planets, the 6 planets with an initial $f_{\rm core} \leq$ 15\% are unable to maintain a stable planetary structure at the onset of our simulation. 
As a result, the right panel of Figure \ref{fig:3} only displays the evolution tracks of the four planets with initial $f_{\rm core}$ values of 20\%, 30\%, 40\% and 50\%.
For the planet with an initial $f_{\rm core}$ of 20\%, its radius is significantly inflated due to the combined effects of intrinsic luminosity and intense incoming stellar radiation.
This leads to a substantial mass-loss rate via hydrodynamic atmospheric evaporation, causing it to lose its entire gaseous envelope in $\sim$ 100 million years.
The planet  with an initial $f_{\rm core}$ of 30\% retains only a minor fraction of its initial gaseous envelope, ultimately settling in the hot Neptune region of the $a$-$R$ parameter space.
In contrast, the two planets with higher initial $f_{\rm core}$ (40\% and 50\%, planets 01 and 02 in Table \ref{tab:2})  retain proportionally more gas and evolve into hot Saturn analogues by the end of the simulation.

Similar to the analysis in Section \ref{sec:3.2}, the final outcome of the radius evolution of a Saturn-mass short-period gaseous planet is primarily determined by the strength of the atmospheric evaporation process.
However, in this case, the main factor controlling the strength of the atmospheric evaporation is the initial $f_{\rm core}$.
Among this group of 10 synthetic planets, only the two planets with large $f_{\rm core}$ values (40\% and 50\%) are able to retain measurable envelopes despite intense atmospheric evaporation processes, ultimately evolving into hot Saturn analogues.

\subsection{Hot Saturns around M-dwarf stars}
\label{sec:3.4}

\begin{table*}[]
\centering
\caption{Initial and final properties of hot Saturn analogues emerging from 10,000 synthetic planets orbiting an M1-type star.}
\label{tab:3}
\begin{tabular}{ccccccccccc} 
\toprule
 & & \multicolumn{5}{c}{Initial} & \multicolumn{4}{c}{Final} \\
\cmidrule(lr){3-7} \cmidrule(lr){8-11}
Index & $a$  &  $M_{\rm p}$  &  $f_\mathrm{core}$ &  $L_{\rm p}$  &  $S_\mathrm{inner}$$^{(1)}$ & $S_\mathrm{outer}$$^{(2)}$ &  $M_{\rm p}$  & $f_\mathrm{core}$ &  $L_{\rm p}$  &  $R_{\rm p}$  \\
 &  (AU) &  ($\mathrm M_{\rm Jup}$) &  & ($\mathrm L_{\odot}$)  & ($K_\mathrm{B}/baryon$) &($K_\mathrm{B}/baryon$) & ($\mathrm M_{\rm Jup}$) &  &  ($\mathrm L_{\odot}$) & ($\mathrm R_{\oplus}$) \\ 
 \midrule
 01&  0.03&  0.4&  30\%&  $1.9 \times 10^{-6}$ & 10.39 &10.42 &  0.14&  87.6\%&  $9.34\times 10^{-11}$& 6.64\\
 02&  0.03&  0.4&  30\%&  $1.7 \times 10^{-5}$ & 10.48 &10.48 &  0.14&  86.5\%&  $1.05\times 10^{-10}$& 6.85\\
 03&  0.03&  0.3&  40\%&  $6.3 \times 10^{-7}$ & 10.26 &10.50 &  0.14&  87.2\%&  $9.68 \times 10^{-11}$& 6.71\\
 04& 0.03& 0.3& 40\%& $1.9 \times 10^{-6}$ & 10.57 &10.69 & 0.14& 87.0\%& $9.98 \times 10^{-11}$& 6.76\\
 05& 0.03& 0.3& 50\%& $1.9 \times 10^{-6}$ & 10.46 &10.47 & 0.20& 76.8\%& $3.03 \times 10^{-10}$& 8.27\\\hline
 \multicolumn{11}{l} { \footnotesize (1) $S_\mathrm{inner}$ is the entropy at the core-envelope boundary.} \\
  \multicolumn{11}{l} { \footnotesize (2) $S_\mathrm{outer}$ is the entropy at the planetary radius of $\tau = 2/3$.} \\
 \bottomrule
 \end{tabular}\end{table*}

A key assumption in our atmospheric evaporation model is that all host stars exhibit solar-type properties, with their X-ray and EUV flux following the same temporal evolution as described by \citet{Ribas2005}.
In reality, different types of host stars exhibit distinct spectral energy distributions across various wavelengths. These variations in the stellar spectral energy distribution can alter the strength and dynamics of the atmospheric evaporation process. As a result, the derived planetary properties of hot Saturn analogues, particularly their initial luminosity and entropy, are also affected.
Given that Kepler and TESS observe both solar-type stars and M-dwarfs, hot Saturn analogues around M-dwarfs are expected to exhibit different initial characteristics, as the temporal evolution of X-ray and EUV flux for M-dwarfs differs from that of solar-type stars.
In this section, we performed a grid study on the evolution of short-period gaseous planets around M-dwarfs, employing the same methodology used in the previous grid study of planets orbiting solar-type stars.
Specifically, we simulated the long-term evolution of 10,000 synthetic gaseous planets around an M1-type star with a mass of 0.5 $M_{\odot}$. 
The initial properties of these simulated planets were identical to those listed in Table \ref{tab:1}. 
The only difference in this grid study lies in the host star’s mass and the temporal evolution of its X-ray and EUV flux, which are set according to the spectral characteristics of M-dwarfs.

To estimate the X-ray flux of our modeled M1-type host star, we employ the activity-age correlation for M0-2 stars as presented in Equation 1 and Table 1 of \citet{Engle2024}.
Since the X-ray activity data from \citet{Engle2024} spans 5–120  ${\rm \AA}$, but only the 5–10   ${\rm \AA}$ range significantly contributes to heating \citep{Owen2012}, we estimate the X-ray flux between 5–20 ${\rm \AA}$ by applying the wavelength-dependent flux proportions from \citet{Ribas2005} (Table 5).
This adjusted flux value is subsequently used to calculate the X-ray-driven mass-loss rate.
To model  the temporal evolution of EUV flux in M-dwarfs, we apply Equations 1 and 2 with the M-star specific coefficients from \citet{Lecavelier2007}. 
Following this, the temporal evolution of the synthetic planets and their mass-loss rates are determined, depending on whether their atmospheric evaporation processes are X-ray-driven or EUV-driven, as described in Section \ref{model:2.3}.

Table \ref{tab:3} presents the hot Saturn analogues that emerged from the grid study of planetary evolution around an M1-type star.
In this case, five hot Saturn analogues were formed. 
Similar to the grid study around solar-type stars, all five hot Saturn analogues initially possess a substantial $f_\mathrm{core}$ ranging from 30\% to 50\%. Over the course of their 1 Gyr evolution, they experience a vigorous atmospheric evaporation process, losing most of their initial gaseous envelopes and ultimately achieving a final $f_\mathrm{core}$ greater than 75\%.
A notable difference from the case of solar-type stars is that the initial total planetary masses of these five hot Saturn analogues range from 0.3 to 0.4 $M_{\rm Jup}$.
In the case of planets orbiting M-dwarfs, we do not observe hot Saturn analogues with an initial mass of 0.5 $M_{\rm Jup}$.
The reason for this is that the atmospheric evaporation process around M-dwarfs is weaker than that around solar-type stars, due to the lower stellar flux. As a result, planets with an initial mass of 0.5 $M_{\rm Jup}$ only lose a limited portion of their gaseous envelopes and ultimately evolve into hot Jupiter planets.
The initial luminosities of the hot Saturn analogues formed around M-dwarf stars range from $6.3 \times 10^{-7}$ to $1.7 \times 10^{-5}$ $L_{\odot}$, with initial entropies around  10.5 $K_\mathrm{B}/baryon$.
Compared to hot Saturn analogues around solar-type stars, they exhibit relatively higher initial $L_{\rm p}$  for a given total planetary mass.

The results show that although hot Saturn analogues around different stellar types exhibit variations in initial luminosity and total planetary mass—stemming from different evaporation strengths due to their distinct stellar flux histories—they share key characteristics that ultimately lead them to reside in the hot Saturn desert region.
These characteristics include a sub-Jovian total mass less than 0.5 $M_{\rm Jup}$, a substantial $f_{\rm core}$ between 30\% and 50\%, and an initial $L_{\rm p}$ within a narrow, permissible range.
These properties cause them to undergo intense atmospheric evaporation processes, stripping away most of their initial gaseous envelope, resulting in a final $f_{\rm core}$ greater than 75\%, and ultimately placing them in the hot Saturn desert region.

\section{DISCUSSION and SUMMARY}

\label{sec:discussion}

The evolution of gaseous planets in the hot Saturn desert region of the $a$-$R$ space is strongly influenced by atmospheric evaporation due to the intense stellar flux at short-period orbits.
Given this context, the scarcity of hot Saturns in the observed exoplanet population prompts an important question: Why are there so few Saturn-sized exoplanets in short-period orbits after undergoing a long evolutionary process involving atmospheric evaporation?
Our grid simulations of the evolution of synthetic planets around solar-type and M-dwarf stars reveal the significant challenges in forming hot Saturn analogues.
If a short-period gaseous planet has a sufficiently strong gravitational potential, such as an initial Jupiter-mass planet, it will experience limited mass loss due to atmospheric evaporation throughout its evolution, ultimately becoming a hot Jupiter analogue.
Conversely, if a short-period planet’s gravitational binding energy is insufficient to counteract atmospheric evaporation—such as in the case of an initial Neptune-mass planet, or an initial Saturn-mass planet with a low $f_{\rm core}$—it is more likely to lose its entire gaseous envelope and end up as a bare rocky core, appearing as a super-Earth in the $a$-$R$ space.
Only sub-Jovian planets with initial masses less than 0.5 $M_{\rm Jup}$ and a substantial $f_{\rm core}$ between 30\% and 50\% can maintain the delicate balance required for the formation of a hot Saturn analogue.
These planets must lose most of their initial gaseous envelope, resulting in a final $f_{\rm core}$ greater than 75\%. 
This unique combination of a large core and a remnant gaseous envelope less than $\sim$ 20\% is the only planetary configuration capable of persisting within the hot Saturn desert region.
This is particularly reasonable, considering that the observed radii of short-period exoplanets with masses ranging from 0.5 to 5.0  $M_{\rm Jup}$ are generally larger than 1.0  $R_{\rm Jup}$.

In addition to the initial planetary mass and the initial $f_{\rm core}$, the initial intrinsic luminosity of a planet also plays a crucial  role in determining the strength of the atmospheric evaporation process and, consequently, the final planetary radius.
Section \ref{sec:3.2} demonstrates  that the initial luminosity of a short-period, Saturn-mass planet is an important factor in determining whether the planet will eventually evolve into a hot Saturn.
For a planet with an initial mass of $0.5$ $M_\mathrm{J}$ and an initial $f_{\rm core}$ of 40\% at 0.03 AU around a solar-type star, it will evolve into a hot Jupiter only at an initial intrinsic luminosity of $1.9 \times 10^{-6} L_{\odot}$.
In contrast, it will not evolve into a hot Jupiter if its initial luminosity is $\geq$  $5.7\times 10^{-6} L_{\odot}$ or  $\leq$  $6.3 \times 10^{-7} L_{\odot}$.
Theoretical studies indicate that the luminosity of a Jupiter-mass planet is approximately $10^{-5}-10^{-4} L_{\odot}$ at 1 million years and around  $10^{-8} L_{\odot}$  at 1 billion-year \citep{Marleau2014}. 
Given that hot Saturns are a rare phenomenon among exoplanets, the initial luminosities and entropies of our synthetic hot Saturn analogues listed in Table \ref{tab:2} can be regarded as implausible initial conditions for Saturn-mass planets with a large rocky core.

The formation of our synthetic hot Saturn analogues is closely linked to the atmospheric evaporation process. Consequently, the details of the atmospheric evaporation model adopted will also affect the results of our simulations. 
In this work, we use the atmospheric evaporation model of \citet{Jin2018} and do not consider the  effects of varying the values of critical parameters, such as the escape efficiency $\epsilon$ in the energy-limited evaporation model. 
Choosing an alternative value for the parameter $\epsilon$ will result in different derived initial parameters that can form hot Saturn analogues, affecting the results in a manner similar to the hot Saturn analogues formed around M-dwarfs as shown in Section \ref{sec:3.4}.
Presumably, the expected variations in the results will be limited, considering that doubling the value of $\epsilon$ has only a minimal impact on the final evolutionary outcome of gaseous planets \citep{Jin2014}.
Incorporating variations in the atmospheric evaporation modelling will introduce an additional degree of freedom to our analysis, which will be explored in more detail in our future work.

Another noteworthy point is that this paper defines hot Saturn analogues as synthetic planets with a radius between 6 and 8.5 $R_{\oplus}$. 
This range is slightly smaller than Saturn's actual radius and could be even smaller than of a Saturn-mass planet under intense stellar radiation at a short-period orbit.
This is because, although our planetary structure model for short-period gas giants includes the effects of strong stellar radiation on planetary radii \citep{Guillot2010}, it still cannot accurately match the observed larger radii of short-period gas giants, a discrepancy known as radius anomaly problem in hot Jupiters \citep{Sestovic2018,Jin2025}.
A smaller planetary radius corresponds to less received stellar flux, meaning that the shortcoming of our planetary structure model results in a lower calculated mass-loss rate due to atmospheric evaporation.
However, since the long-term evolution of gas giants is not particularly sensitive to the limited changes in mass-loss rate of atmospheric evaporation \citep{Jin2014}, the derived initial physical parameters of synthetic hot Saturn analogues will remain largely unchanged.
The main consequence of this shortcoming in our model is that the hot Saturn desert we are discussing here is situated between 6 and 8.5 $R_{\oplus}$, whereas in reality the hot Saturn desert would correspond to a slightly larger radius interval, due to the fact that our planetary structure model produces relatively smaller planetary radii.

The main result of this work is that if a short-period planet is to eventually reside in the hot Saturn desert region after 1 billion years of evolution, it must have initial physical parameters—immediately following the dissipation of the gas disk—corresponding to a total mass of less than 0.5 or 0.4  $M_{\rm Jup}$, an initial $f_{\rm core}$ larger than 30\%, and an intrinsic  $L_{\rm p}$ on the order of  $10^{-6}$ $L_{\odot}$ or less, orbiting a solar-type or M-type star.
However, hot Saturn planets are extremely rare in the exoplanet population.
This implies that planets with the initial parameters outlined above are difficult to form, or may only form in orbits that are distant from their host stars and are unable to migrate to short-period orbits while a gaseous disk is still present. 
Considering that planet formation theories suggest that the initial physical properties of our hot Saturn analogues (i.e., Saturn-mass planets with large rocky cores in short-period orbits just after the dissipation of a gas disk) are unfavorable according to the core accretion scenario \citep{Liu2020}, the existence of a hot Saturn desert in the $a$-$R$ space actually lends support to the core accretion scenario.
Another useful finding of this work is that for a short-period gaseous planet that ends up in the hot Saturn desert region after a long-term evolution, such as the recently identified TOI-1194 b \citep{Wang2024}, it would be expected to have a substantial$f_{\rm core}$  based on the properties of our synthetic hot Saturn analogues.

\vspace{0.9cm}
We are grateful to an anonymous referee for providing constructive comments that have improved this paper.
This work is supported by National Natural Science Foundation of China (Grant No. 11973094, 12103003), the incubation program for recruited talents (2023GFXK153) and the doctoral start-up funds from Anhui Normal University. 

\software{\texttt{gnuplot} \citep{Williams2022}}

\bibliographystyle{aasjournal}

\end{document}